
%

\hsize=6.0truein
\vsize=8.5truein
\voffset=0.25truein
\hoffset=0.1875truein
\tolerance=1000
\hyphenpenalty=500
\def\monthintext{\ifcase\month\or January\or February\or
   March\or April\or May\or June\or July\or August\or
   September\or October\or November\or December\fi}


\font\tenrm=cmr10 scaled \magstep1   \font\tenbf=cmbx10 scaled \magstep1
\font\sevenrm=cmr7 scaled \magstep1  
\font\fiverm=cmr5 scaled \magstep1   

\font\teni=cmmi10 scaled \magstep1   \font\tensy=cmsy10 scaled \magstep1
\font\seveni=cmmi7 scaled \magstep1  \font\sevensy=cmsy7 scaled \magstep1
\font\fivei=cmmi5 scaled \magstep1   \font\fivesy=cmsy5 scaled \magstep1

\font\tentt=cmtt10 scaled \magstep1
\font\tenit=cmti10 scaled \magstep1
\font\tensl=cmsl10 scaled \magstep1

\def\twelvepoint{\def\rm{\fam0\tenrm}
   \textfont0=\tenrm \scriptfont0=\sevenrm \scriptscriptfont0=\fiverm
   \textfont1=\teni  \scriptfont1=\seveni  \scriptscriptfont1=\fivei
   \textfont2=\tensy \scriptfont2=\sevensy \scriptscriptfont2=\fivesy
   \textfont\itfam=\tenit \def\it{\fam\itfam\tenit}
   \textfont\ttfam=\tentt \def\tt{\fam\ttfam\tentt}
   \textfont\bffam=\tenbf \def\bf{\fam\bffam\tenbf}
   \textfont\slfam=\tensl \def\sl{\fam\slfam\tensl} \rm
   \hfuzz=1pt\vfuzz=1pt
   \setbox\strutbox=\hbox{\vrule height 10.2pt depth 4.2pt width 0pt}
   \parindent=24pt\parskip=1.2pt plus 1.2pt
   \topskip=12pt\maxdepth=4.8pt\jot=3.6pt
   \normalbaselineskip=14.4pt\normallineskip=1.2pt
   \normallineskiplimit=0pt\normalbaselines
   \abovedisplayskip=13pt plus 3.6pt minus 5.8pt
   \belowdisplayskip=13pt plus 3.6pt minus 5.8pt
   \abovedisplayshortskip=-1.4pt plus 3.6pt
   \belowdisplayshortskip=13pt plus 3.6pt minus 3.6pt
   \topskip=12pt \splittopskip=12pt
   \scriptspace=0.6pt\nulldelimiterspace=1.44pt\delimitershortfall=6pt
   \thinmuskip=3.6mu\medmuskip=3.6mu plus 1.2mu minus 1.2mu
   \thickmuskip=4mu plus 2mu minus 1mu
   \smallskipamount=3.6pt plus 1.2pt minus 1.2pt
   \medskipamount=7.2pt plus 2.4pt minus 2.4pt
   \bigskipamount=14.4pt plus 4.8pt minus 4.8pt}

\twelvepoint



\font\titlerm=cmr10 scaled \magstep3
\font\titlerms=cmr10 scaled \magstep1 
\font\titlei=cmmi10 scaled \magstep3  
\font\titleis=cmmi10 scaled \magstep1 
\font\titlesy=cmsy10 scaled \magstep3      
\font\titlesys=cmsy10 scaled \magstep1  
\font\titleit=cmti10 scaled \magstep3 
\skewchar\titlei='177 \skewchar\titleis='177 
\skewchar\titlesy='60 \skewchar\titlesys='60 

\def\titlefont{\def\rm{\fam0\titlerm}
   \textfont0=\titlerm \scriptfont0=\titlerms 
   \textfont1=\titlei  \scriptfont1=\titleis  
   \textfont2=\titlesy \scriptfont2=\titlesys 
   \textfont\itfam=\titleit \def\it{\fam\itfam\titleit} \rm}


\def\preprint#1{\baselineskip=19pt plus 0.2pt minus 0.2pt \pageno=0
   \begingroup
   \nopagenumbers\parindent=0pt\baselineskip=14.4pt\rightline{#1}}
\def\title#1{
   \vskip 0.9in plus 0.45in
   \centerline{\titlefont #1}}
\def\secondtitle#1{}
\def\author#1#2#3{\vskip 0.9in plus 0.45in
   \centerline{{\bf #1}\myfoot{#2}{#3}}\vskip 0.12in plus 0.02in}
\def\secondauthor#1#2#3{}
\def\addressline#1{\centerline{#1}}
\def\abstract{\vskip 0.7in plus 0.35in
   \centerline{\bf Abstract}
   \smallskip}
\def\finishtitlepage#1{\vskip 0.8in plus 0.4in
   \leftline{#1}\supereject\endgroup}

\def\date#1{\finishtitlepage{#1}}

\def\nolabels{\def\eqnlabel##1{}\def\eqlabel##1{}\def\figlabel##1{}%
   \def\reflabel##1{}}
\def\writelabels{\def\eqnlabel##1{%
   {\escapechar=` \hfill\rlap{\hskip.11in\string##1}}}%
   \def\eqlabel##1{{\escapechar=` \rlap{\hskip.11in\string##1}}}%
   \def\figlabel##1{\noexpand\llap{\string\string\string##1\hskip.66in}}%
   \def\reflabel##1{\noexpand\llap{\string\string\string##1\hskip.37in}}}
\nolabels


\global\newcount\secno \global\secno=0
\global\newcount\meqno \global\meqno=1

\def\newsec#1{\global\advance\secno by1
   \xdef\secsym{\the\secno.}
   \global\meqno=1\bigbreak\medskip
   \noindent{\bf\the\secno. #1}\par\nobreak\smallskip\nobreak\noindent}
\xdef\secsym{}

\def\appendix#1#2{\global\meqno=1\xdef\secsym{\hbox{#1.}}\bigbreak\medskip
\noindent{\bf Appendix #1. #2}\par\nobreak\smallskip\nobreak\noindent}


\def\eqnn#1{\xdef #1{(\secsym\the\meqno)}%
   \global\advance\meqno by1\eqnlabel#1}
\def\eqna#1{\xdef #1##1{\hbox{$(\secsym\the\meqno##1)$}}%
   \global\advance\meqno by1\eqnlabel{#1$\{\}$}}
\def\eqn#1#2{\xdef #1{(\secsym\the\meqno)}\global\advance\meqno by1%
   $$#2\eqno#1\eqlabel#1$$}


\def\myfoot#1#2{{\baselineskip=14.4pt plus 0.3pt\footnote{#1}{#2}}}
\global\newcount\ftno \global\ftno=1
\def\foot#1{{\baselineskip=14.4pt plus 0.3pt\footnote{$^{\the\ftno}$}{#1}}%
   \global\advance\ftno by1}


\global\newcount\refno \global\refno=1
\newwrite\rfile

\def\ref{[\the\refno]\nref}
\def\nref#1{\xdef#1{[\the\refno]}\ifnum\refno=1\immediate
   \openout\rfile=refs.tmp\fi\global\advance\refno by1\chardef\wfile=\rfile
   \immediate\write\rfile{\noexpand\item{#1\ }\reflabel{#1}\pctsign}\findarg}
\def\findarg#1#{\begingroup\obeylines\newlinechar=`\^^M\passarg}
   {\obeylines\gdef\passarg#1{\writeline\relax #1^^M\hbox{}^^M}%
   \gdef\writeline#1^^M{\expandafter\toks0\expandafter{\striprelax #1}%
   \edef\next{\the\toks0}\ifx\next\null\let\next=\endgroup\else\ifx\next\empty%

\else\immediate\write\wfile{\the\toks0}\fi\let\next=\writeline\fi\next\relax}}
   {\catcode`\%=12\xdef\pctsign{
\def\striprelax#1{}

\def\semi{;\hfil\break}
\def\addref#1{\immediate\write\rfile{\noexpand\item{}#1}} 

\def\listrefs{\vfill\eject\immediate\closeout\rfile
   \centerline{{\bf References}}\bigskip{\frenchspacing%
   \catcode`\@=11\escapechar=` %
   \input refs.tmp\vfill\eject}\nonfrenchspacing}

\def\startrefs#1{\immediate\openout\rfile=refs.tmp\refno=#1}


\global\newcount\figno \global\figno=1
\newwrite\ffile
\def\fig{\the\figno\nfig}
\def\nfig#1{\xdef#1{\the\figno}\ifnum\figno=1\immediate
   \openout\ffile=figs.tmp\fi\global\advance\figno by1\chardef\wfile=\ffile
   \immediate\write\ffile{\medskip\noexpand\item{Fig.\ #1:\ }%
   \figlabel{#1}\pctsign}\findarg}

\def\listfigs{\vfill\eject\immediate\closeout\ffile{\parindent48pt
   \baselineskip16.8pt\centerline{{\bf Figure Captions}}\medskip
   \escapechar=` \input figs.tmp\vfill\eject}}


\def\letter{\raggedright\parindent=0pt}
\def\endmode{}
\def\longindent{\parindent=3.25truein\obeylines\parskip=0pt}
\def\letterhead{\null\vfil\begingroup
   \parindent=3.25truein\obeylines
   \def\endmode{\medskip\endgroup}}

\def\sendingaddress{\endmode\begingroup
   \parindent=0pt\obeylines\def\endmode{\medskip\endgroup}}

\def\salutation{\endmode\begingroup
   \parindent=0pt\obeylines\def\endmode{\medskip\endgroup}}

\def\body{\endmode\begingroup\parskip=\smallskipamount
   \def\endmode{\medskip\endgroup}}

\def\closing{\endmode\begingroup\longindent
   \def\endmode{\endgroup}}

\def\signed{\endmode\begingroup\longindent\vskip0.8truein
   \def\endmode{\endgroup}}

\def\endofletter{\endmode \ifnum\pageno=1 \nopagenumbers\fi
   \vfil\vfil\eject\end}


\def\noblackbox{\overfullrule=0pt}
\def\inv{^{\raise.18ex\hbox{${\scriptscriptstyle -}$}\kern-.06em 1}}
\def\dup{^{\vphantom{1}}}
\def\Dsl{\,\raise.18ex\hbox{/}\mkern-16.2mu D} 
\def\dsl{\raise.18ex\hbox{/}\kern-.68em\partial}
\def\slash#1{\raise.18ex\hbox{/}\kern-.68em #1}
\def\lspace{}
\def\lbspace{}
\def\boxeqn#1{\vcenter{\vbox{\hrule\hbox{\vrule\kern3.6pt\vbox{\kern3.6pt
   \hbox{${\displaystyle #1}$}\kern3.6pt}\kern3.6pt\vrule}\hrule}}}
\def\mbox#1#2{\vcenter{\hrule \hbox{\vrule height#2.4in
   \kern#1.2in \vrule} \hrule}}  
\def\bar{\overline}
\def\e#1{{\rm e}^{\textstyle#1}}
\def\del{\partial}
\def\curly#1{{\hbox{{$\cal #1$}}}}
\def\curlyD{\hbox{{$\cal D$}}}
\def\curlyL{\hbox{{$\cal L$}}}
\def\vev#1{\langle #1 \rangle}
\def\psibar{\overline\psi}
\def\lform{\hbox{$\sqcup$}\llap{\hbox{$\sqcap$}}}
\def\darr#1{\raise1.8ex\hbox{$\leftrightarrow$}\mkern-19.8mu #1}
\def\half{{\textstyle{1\over2}}} 
\def\roughly#1{\ \lower1.5ex\hbox{$\sim$}\mkern-22.8mu #1\,}
\def\MSbar{$\bar{{\rm MS}}$}
\hyphenation{di-men-sion di-men-sion-al di-men-sion-al-ly}
%
%
%
\def\draftordate{\date{6/92}}
\def\hf{\half}
\def\nonp{non-perturbative}
\def\hmm{hermitian matrix model}
\def\integ#1#2#3{\int_{#1}^{#2}\!\!\! d#3\ }
\def\O{{\cal O}}
\def\R{{\cal R}}
\def\S{{\cal S}}
\def\G{{\Gamma}}
\def\rline{{\rm I}\!{\rm R}} 
\def\pq{$[P,Q]=1$}
\def\pqq{$[{\tilde P},Q]=Q$ }
\def\nuke{Nucl.Phys.}
\def\pl{Phys.Lett.}
\def\re{\rho_R}
\def\im{\rho_I}
\def\trho{\upsilon}
\def\NP#1{Nucl. Phys.\ {\bf #1}\ }
\def\PL#1{Phys. Lett.\ {\bf #1}\ }
\def\CMP#1{Comm. Math. Phys.\ {\bf #1}\ }
\def\LNC#1{Lett. Nuovo Cimento\ {\bf #1}\ }
\def\NC#1{Nuovo Cimento\ {\bf #1}\ }
\def\CQG#1{Class. Quantum Grav.\ {\bf #1}\ }
\def\PR#1{Phys. Rev\ {\bf #1}\ }
\def\PREP#1{Phys. Rep\ {\bf #1}\ }
\def\PRL#1{Phys. Rev. Lett\ {\bf #1}\ }
\def\JMP#1{J. Math. Phys\ {\bf #1}\ }
\def\SAM#1{Stud. Appl. Math\ {\bf #1}\ }
\def\MPL#1{Mod. Phys. Lett\ {\bf #1}\ }
\def\PA#1{Physica\ {\bf #1}\ }
\def\PP#1{Princeton preprint\ { #1}\ }
\def\RP#1{Rutgers preprint\ { #1}\ }

\def\sec#1{
\bigskip
\noindent{\bf #1}
\bigskip}

\preprint{\vbox{\rightline{PUPT--1325}
\vskip2pt\rightline{SHEP 91/92--19}
\vskip2pt\rightline{G\"oteborg ITP 92--20}
\vskip2pt\rightline{hep-th@xxx/9206060}}}
\vskip -1.5cm
\title{Unitary Matrix Models and 2D Quantum Gravity}
\author{S. Dalley$^1$, C.V. Johnson$^2$, T.R. Morris$^2$
and A. W\"atterstam$^3$}{}{}
\addressline{\it ${}^1$Joseph Henry Laboratories}
\addressline{\it Princeton University}
\addressline{\it Princeton NJ 08544, U.S.A.}
\addressline{and}
\addressline{\it ${}^2$Physics Department}
\addressline{\it The University of Southampton}
\addressline{\it SO9 5NH, U.K.}
\addressline{and}
\addressline{\it ${}^3$Institute for Theoretical Physics}
\addressline{\it Chalmers Institute of Technology}
\addressline{\it S--412 96 G\"oteborg, Sweden}
\vskip -0.5cm

\abstract
%
The KdV and modified KdV integrable hierarchies are shown to be different
descriptions of the same 2D gravitational system -- open-closed string theory.
Non-perturbative solutions of the multi-critical unitary matrix
models map to non-singular
solutions of the `renormalisation group' equation for the string
susceptibility, \pqq .
We also demonstrate that the large N solutions of unitary matrix integrals
in external fields, studied by Gross and
Newman, equal the non-singular pure closed-string solutions of \pqq.

\draftordate

\sec{Introduction.}
In this letter we shall show how double-scaling limits of unitary matrix
models can be interpreted in terms of 2D quantum gravity and its
worldsheet expansions. The modified KdV and KdV integrable hierarchies
are shown to be different descriptions of the same string theory.
Specifically, using the Miura transformation between these
hierarchies, and the inverse transformation which we
explicitly construct, we map the solutions
of the double-scaling
limit of symmetric unitary matrix models
with $C$ flavours of massless `quark' \ref\periwal{V.Periwal
and D.Shevitz,  Phys.Rev.Lett. {\bf 64} (1990)  1326, \NP{B344} (1990) 731.}
\ref\Minahan{J.Minahan, Phys. Lett.
{\bf B268} (1991) 29, Phys.Lett. {\bf B265} (1991) 382.}
to non-perturbative solutions of
open-closed string theory in the $(2,2m-1)$ minimal model backgrounds.
 The open string coupling $\Gamma$,
measured in units of the closed string coupling $\nu$, satisfies
$C = 1/2 \pm \Gamma$ and the boundary cosmological constant is zero.
The map implies that the two physical
$\tau$-functions of the mKdV/KdV hierarchy
\ref\TimH{T.Hollowood, L.Miramontes, C.Nappi and A.Pasquinucci,
\NP{B373} (1992) 247}
equal square roots of partition functions of
open-closed strings in general. As a simple corollary
the  continuum limit $1/N$-expansion
of the original unitary matrix
 model partition function\periwal\ is seen to be
nothing other than world-sheets with even
numbers of boundaries and $\G^2=1/4$.
 The general form of the Virasoro
constraints on the $\tau$-function are explained.
Correlators of local operators can be evaluated
in terms of the flows of the hierarchy in the usual way.

Our analysis also leads to a proof of
some transformation equations between solutions of the string equations
with open string couplings differing by
an integer, generalising a  relation proved by Lukashevich, Fokas and
Ablowitz\ref\transf{N.A. Lukashevich, Diff. Urav. {\bf 7} (1971) 1124\semi
A.S. Fokas and M.J. Ablowitz \JMP23 (1982) 2033.} for solutions of
Painlev\'e II. Although the Miura transformation from the mKdV hierarchy
applies to very general solutions, of
particular interest are the unique real non-singular solutions furnished
by the unitary matrix models themselves \ref\anders{A.  W\"{a}tterstam,
\PL{B263} (1991)  51.}\ref\moo{{\v C}.Crnkovi\'c, M.Douglas and G.Moore,
\NP{B360} (1991) 507.}. These map to  open string generalisations of
the non-singular solutions of the KdV hierarchy\foot{For the
case of coupling  to the general
 $(p,q)$ minimal models see also
ref.\ref\pqmodels{C.V.Johnson,  T.R.Morris and  B.Spence, Southampton preprint
SHEP  90/91--30, Imperial  preprint TP/91--92/01, hepth@xxx9203022.}}\
found in
\ref\multi{S.Dalley,
C.V.Johnson and T.R.Morris, \NP{B368} (1992) 625.}\ref\npqg{S.Dalley,
  C.V.Johnson and T.R.Morris,
\nuke\  {\bf  B368}  (1992)  655.}.
Remarkably, the pure {\sl closed string} partition
functions uncovered in those
papers are almost surely the solutions determined by Gross and
Newman \ref\gni{D.J. Gross  and M.J. Newman,
\PL{266} (1991) 291.}\ref\gnii{D.J. Gross and M.J. Newman,
Princeton   preprint    PUPT-1282, hepth@xxx9112069.} for the simplest
unitary matrix model in an external field  and its formal
multicritical generalisations. Damning evidence of this fact will be
presented in a later section.
The main results are expanded upon in the following section, but first we
conclude this introduction by recalling
some salient points about open-closed string theory and matrix models.

Solutions $u(z,t_{k})$ of the KdV hierarchy\foot{$z$ is essentially the `space'
variable in the KdV equations, the $t_k$ with $k\geq 1$ being generalised time
variables.}\ $\partial_{t_{k}}u=\partial_{z}
\R_{k+1}[u]$ must satisfy a couple of physical requirements if they are to
represent the string susceptibility of a 2D gravity theory. Firstly as a
function of the dimensionful arguments $\{t_{k},z\}$ they must satisfy
the renormalisation
group (RG) equation for invariance under a change of scale \multi\ , given by
\pqq in terms of pseudo-differential operators \npqg\ . After some simple
manipulation one can rewrite it \multi\ :
 \eqn\smiley{u{\cal R}^2-{1\over
 2}{\cal  R}{\cal R}^{''}+{1\over  4}({\cal R}^{'})^2=\G^{2} }
where ${}'\equiv \partial_{z}$, $\R$ is defined in terms of Gel'fand-Dikii
differential polynomials $\R_{k}[u]$ as $\R = \sum_{k=1}(k+1/2)
 t_{k} \R_{k}-z$, and
$\G^{2}$ arises as an integration constant. Secondly $u$ must
have an asymptotic expansion in $z$ as
$z \rightarrow \infty$ which matches that obtained from the
perturbative genus expansion
of 2D gravity coupled to some matter theory.
 For the (flows between the) $(2,2m-1)$ minimal models coupled to
gravity this means that for closed strings $u$ must satisfy $\R = 0$ to all
orders in the asymptotic expansion. Eqn.\smiley\ has such solutions provided
that $\G=0$. Indeed, either $\R=0$ exactly \ref\orig{ E.Br\'{e}zin  and
V.Kazakov,
Phys.Lett. {\bf B236}
(1990)  144\semi  M.Douglas  and  S.H.Shenker,  Nucl.Phys.  {\bf  B335} (1990)
635\semi  D.J.Gross and  A.A.Migdal, Phys.Rev.Lett.  {\bf 64}  (1990) 127,
  Nucl.    Phys.   {\bf   B}340   (1990)   333.} yielding solutions which are
complex \ref\rus{P.G.Silvestrov and
A.S.Yelkhovskii, \PL{B251} (1990) 525.}\ref\david{F.David,
Mod. Phys. Lett. {\bf A5} (1990) 1019,
Nucl.Phys. {\bf B348} (1991) 507.}
or have singularities somewhere on the flow
\ref\noflow{M.R.Douglas,
N.Seiberg and S.Shenker, \PL{B244}\ (1990) 381.}, or one
may choose the solution
with unique real non-singular flow studied in
refs.\multi\npqg\ref\flows{C.V.Johnson,    T.R.Morris   and
A.W\"atterstam,  Southampton  preprint  SHEP  91/92--25  and G\"oteborg ITP
92--21, hepth@xxx9205056.} (referred to as DJM solution hereafter).
$\G$ plays the role of an
open string coupling as follows by comparing eqn.\smiley\ with the string
equation found by Kostov \ref\kostov{I.K.Kostov, \PL{B238} (1990) 181.},
obtained by performing the double-scaling limit of matrix models with a
term ${\G \over N}{\rm Tr}\log{(1+\phi^2)}$
in the potential to generate random
holes \ref\kazak{V.A.Kazakov, \PL{B237} (1990) 212.}; each surface is weighted
by $\G^{\# {\rm holes}}$. In fact Kostov's solutions were restricted to those
for which
\eqn\ivan{\R=- 2\G {\rm diag}\{(-\partial_{z}^{2}+u)^{-1}\}}
these obeying $\R=0$ in the $\G \to 0$ limit. Hence
eqn.\smiley\ is more general, encompassing the
open string generalisations of the DJM closed string solutions, the previous
relation \ivan\ then holding as a matching condition
in the sense of an asymptotic expansion as $z\to\infty$,
but being
violated at the non-perturbative level. Note that
the renormalised boundary cosmological constant $\rho$ which one may assign
to the holes, determined by the position of the branch point of the logarithm,
has been set to zero in \smiley\ since this is the case which will interest us
in what follows. We shall discuss $\rho \neq 0$ briefly later, in the context
of the Virasoro constraints.

Introducing the closed string coupling $\nu$ (renormalised $1/N$)
 into the string equations \smiley\ and \ivan\
by the rescaling $t_{k} \rightarrow t_{k}/\nu, z \rightarrow z/\nu$,
the asymptotic solution is
a series in $\nu$ which at the $m$th critical point takes the form:
\eqn\posex{u= z^{1/m}\sum_{g,h=0}^{\infty} A_{gh}
{\nu^{2g+h}\G^{h} \over z^{(2+1/m)(g+h/2)}}}
$A_{gh}$ is determined in \ivan\ once $A_{00}$
 has been fixed,
and determined in \smiley\ once $A_{00}$ and the sign of
$A_{01}$ have been fixed. For the sake of clarity let us rescale such that
$(m+1/2)t_m\R_m=u^m+\cdots$, then in the
 case $z \rightarrow +\infty$  the sphere
term is required to be $A_{00}=1$. Using \ivan\ then fixes the disc term to
$A_{01}=-1/m$.
We shall also need later the real
$z \rightarrow -\infty$ expansion particular to the DJM solution,
obtained by taking the new
possibility allowed by eqn.\smiley, $A_{00}=0$:
\eqn\negex{u_{\rm DJM}(z \rightarrow -\infty) =
{\nu^2\over z^{2}}\left(\G^2-{1\over4}\right)
\sum_{r=0}^{\infty}\sum_{s=0}^{mr} B_{rs}
{\nu^{2mr}\G^{2s}\over z^{(2m+1)r}}\ \ .}
Substituting \negex\ in \smiley , the torus and cylinder terms
are readily  seen to be as above with $B_{00}=1$.
The general form then follows from uniqueness together with
the symmetry of eqn.\smiley\
(after introducing $\nu$) under $\nu^{2} \rightarrow j\nu^{2}$, $u \rightarrow
ju$ where $j$ is any $m$th root of one.

\sec{String Equations and the Miura Map.}
Solutions $v(z,t_{k})$ of the  modified KdV hierarchy are defined
by the flow equations $\partial_{t_{k}}
v = \hf \partial_{z} \S_{k}[v]$ where $\S_{k}\equiv \hf
\R_{k}^{\prime}[v^{2}+v'] -
v\R_{k}[v^{2} +v']$. Here $v^2$ plays
the role of string susceptibility, and its analogous
 RG equation  is just the differential of the unitary
matrix model string equation \multi\
given by the flowing version of the Painlev\'e II hierarchy:
\eqn\pII{\sum_{k=1}^\infty
(k+{1\over 2})t_k \S_k[v(z)]+zv(z)=C}
($C$ is again an integration constant). This equation
with $C=0$ was first found by Periwal and Shevitz \periwal\
by taking the double-scaling limit of a unitary matrix model with partition
function:
\eqn\unit{Z = \int {\cal D}U {\rm exp}(N{\rm Tr}\sum g_{p}(U+U^{\dagger})^{p})}
The eigenvalues of $U$ lie on the unit circle and the critical point occurs
when the support of their large $N$ density function just covers $2\pi$. The
two ends of this support meet to produce $2m$ zeros when the potential is
tuned to the $m$th critical point. The significance of $C\neq0$ was later
understood by Minahan \Minahan\ who derived eqn.\pII\ by adding a term
$\pm {2C\over N} {\rm Tr}\log{|1+U|}$ to the potential,
which may be obtained by integrating
out $C$ flavours of bosonic or fermionic `quarks'.

We will now display a one to one map between solutions of \pII\
with  $C=1/2\pm {\Gamma}$ and solutions of \smiley\ .
This map is the {\sl Miura  transformation} $u=v^2+v'$ familiar in the theory
of  the KdV and mKdV
equations\foot{The authors of ref.\TimH\ used the Miura map in the
case
$\R=0$ $(\G=0)$ to obtain the Painlev\'e II hierarchy, but incorrectly
dropped the constant $C=1/2$.}.
Furthermore we will define a one to one transformation
between solutions
$v_{C}(z)$ of \pII\ and solutions of \pII\ with  unit change
in the boundary coupling: $v_{C\pm1}(z)$.
 Equivalently between $u_{\Gamma}(z)$
and $u_{|\Gamma\pm1|}(z)$. To avoid confusion we need to state
carefully our conventions. The maps apply to {\sl general}
solutions $u_{\Gamma}$ of \smiley\ and $v_C$ of \pII ,
not
only those obeying the physical boundary conditions discussed
below eqn.\ivan\ and below (13). The value of the
boundary coupling is used to label any one of the set
of solutions with that coupling, and different labels
may refer to {\sl different} functions of the two
arguments e.g. setting $v_\pm(z,C\pm1)=v_{C\pm1}(z)$
we have $v_+(z,A)\ne v_-(z,A)$ in general. Finally
the label $\G$ for solutions $u_\G$ of \smiley\ will
always be taken to be non-negative; the actual sign
of the open string coupling in the physical solutions
is fixed by reference to \ivan\ and will be discussed
later.

Let $u(z)$ be a solution to \smiley\ and define
\eqn\XXX{
X_\pm[u,v] =  \half \R^{'}[u]\mp  {\Gamma} -v(z)\R[u]\ \ .
}
Then $X_\pm[u,v]\equiv0$ defines a function $v(z)$ (possibly
with singularities):
\eqn\inverse{v={\half \R^{'}[u]\mp  {\Gamma} \over \R[u]} \ \ .}
Note that one can define $v$ for the $\G \rightarrow 0$ limit of
Kostov's solutions by using \ivan\ to cancel $\G$ first.
Now
\eqn\proof{0=X_\pm^2\pm
2{\Gamma}X_\pm-\R[u] X_\pm^{'}=(v^2+v'){\cal R}^2[u]-{1\over
2}{\cal  R}[u]{\cal  R}^{''}[u]+{1\over  4}({\cal  R}^{'}[u])^2
-{\Gamma}^2\ \ .}
 By comparison
with \smiley\ we deduce that the inverse transformation is $u=v^2+v'$,
which on substituting in \inverse\ and rearranging gives eqn.\pII\
with $C=1/2\pm\Gamma$. On the other
hand given a solution $v$ of eqn.\pII\
with $C=1/2\pm\Gamma$, and defining
a function $u=v^2+v'$ we have $X_\pm[u,v]=0$ by rearrangement,
and hence \inverse\ is the inverse transformation and
$u$ satisfies \proof\ which is \smiley .

Thus we have constructed a one to one correspondence between
solutions $v_{C}$ of \pII\ with $C=1/2\pm\Gamma$
and $u_{\Gamma}$ of \smiley\ . Since this is true for
both choices of $C$ it follows from
the Miura map that there is a one to one correspondence
between solutions $v_{C}$ and
solutions $v_{1-C}$ such that
$$v_{C}^2+v_{C}^{\prime}=v_{1-C}^2+v_{1-C}^{\prime}$$
 and by using \inverse\
\eqn\flip{v_{1-C}=v_{C}+{ 2C-1\over\R[v_{C}^{2}+
v_{C}^{\prime}]}\ \ ,}
It is easy to see that this `flip' transformation  changes
the sign on  boundary conditions for $v$ given by $v\to\pm
z^{1/2m}$ as $z\to\infty$. On the other hand \pII\ is odd
in $v$ \periwal\ .
 One can see this for example by induction,  using the
linear recurrence relation in ${\cal R}_m$ to prove
$$v\S'_{m+1}-v'\S_{m+1}={1\over4}v\S_m'''-{1\over4}v'S''_m
-v^3\S_m'$$
and noting that since the $\R_m[u]$ contains no $u$-independent
terms, $\S_m$ can contain no terms linear in $v$.
Thus, given
a solution $v_{C}$ we can construct a solution $v_{-C}$
by
\eqn\slip{v_{-C}=-v_{C}\ \ .}
 Combining this with \flip\ we obtain the
following one to one correspondence between solutions which
in particular preserves the above
boundary conditions:
\eqn\trans{v_{C\pm1}=-v_{C}-{ 2C\pm1\over\R[v_{C}^{2}\mp
 v_{C}^{\prime}]}\ \ .}
By using the Miura map we obtain similarly a one to one
correspondence between solutions $u_{\Gamma}(z)$ and
$u_{{\tilde \Gamma}}(z)$ when
$\Gamma\pm{\tilde \Gamma}$ is an integer.
It is clear that there are many properties that can
be deduced about the solutions of \smiley\ and \pII\ from
 these equations.
In particular, generalising ref.\transf , if one starts with
 $v_{0}(z)\equiv0$ one can generate a series of rational
solutions to \pII\ and \smiley\
 with $\Gamma-1/2$ or $C$ an
integer, and finitely many $t_k\ne 0$.

Returning to `physics', if we now re-introduce $\nu$ into \smiley\ and
\pII\ again via $t_{k} \rightarrow t_{k}/\nu$, $z \rightarrow z/\nu$,
one finds at the $m$th critical point that, when it has an asymptotic
expansion, $v$ is of the form
\eqn\vasy{v=\pm z^{1/2m}\sum_{p,q=0}^{\infty} C_{pq} {\nu^{2p+q} C^{q}\over
z^{(2+1/m)(p+q/2)}}\ \ .}
With $t_m$ set as below \posex , $|C_{00}|=1$ or
$C_{00}=0$. In fact the asymptotics $v\to\pm z^{1/2m}$
($v\to0$)
as $z\to+\infty$ ($z\to-\infty$) are the ones required by
unitary matrix models\periwal\Minahan . If  $C>0$ the sign choice
corresponds to $C$ bosonic (fermionic)
quarks\Minahan . ($C<0$ can be flipped to these by \slip\
and in the original $C=0$ case\periwal\ the sign is clearly
irrelevant). For either sign, a simple calculation reveals that
$C_{01}=-1/(2m)$ as $z\to+\infty$, while the leading
asymptotic is $v=C\nu/z$ as $z\to -\infty$. Using the Muira map
($u=v^{2}+\nu\partial_{z}v$) on these low
orders of perturbation theory one readily finds
that the asymptotics map onto those of the DJM
solutions as in eqns.\posex\negex . Indeed this also
 determines the sign of the open
string coupling to be $\G=\pm(C-1/2)$, again the
sign choice being that of the $z\to+\infty$ $v$-asymptotic.
This may be used to determine the effect on  $\G$ of the
maps in \flip\---\trans ; the effect in all cases is to
increase or decrease $\G$ by 1, thus generating one to
one maps between solutions with open string couplings
differing by integers. 

Of particular interest
are the known non-singular solutions.
When $C=0$ the unitary matrix model \unit\
provides a {\sl unique} non-singular flow
 between the critical points \anders\moo\
having the plus-branch boundary conditions at the $m$th critical point.
Trivially by \slip\ there are such solutions for
the minus branch also.
Thus these map under Muira onto DJM-type solutions with
non-singular flows, for open string couplings $\G=1/2$
and $\G=-1/2$. Incidentally for $z\to-\infty$ \negex\ implies that $u$
is zero to all orders in the asymptotic expansion, consistent
with the $v$--solutions since they also have this behaviour.
On the other hand the {\sl unique} non-singular flows of the DJM
solutions\flows\ for $\G=0$ map onto the
fermion branch\foot{The sign
follows from \inverse\ and by inspection of the $m=1$ numerical
 solution.}\ $v$--solutions
with $C=1/2$. If these are non-singular, which we expect, then
they are also the unique solutions with this property.

Generally, there is  to our knowledge  no thorough study of the
 existence
and uniqueness of
non-singular solutions  of  the  Painlev\'e~II  hierarchy with $C\ne0$ in  the
literature, nor for eqn.\smiley\  when $\G \neq 0$.
Employing the  techniques used for  the $C=0$ case \anders\ref\flas{H.Flaschka
  and
A.Newell, \CMP{76} (1980) 65.} has not yet borne fruit for us.
Minahan argued \Minahan\ that  there should be a unique non-singular real
solution for $C>0$ fermionic quarks. These
would map smoothly to real non-singular DJM-type solutions
of open-closed string theory with $\G < 1/2$. Due to \slip\
there would also be a $C<0$ real pole-free solution with
 positive-branch asymptotic, giving $\G < -1/2$. On the
other hand similar arguments suggest that there
may be unique real non-singular $u$-solutions for $\G<0$.

This two-fold branching when $\G \neq 0$ is also reflected
in the $\tau$-functions.
As noted in ref.\TimH , there are for the joint KdV and mKdV system
related
by the Miura map, two tau-functions $\tau_{0}$ and $\tau_{1}$ arising from
the two basic representations of the Ka{\v c}-Moody algebra $A_{1}^{(1)}$.
They satisfy
\eqn\taufn{v^{2} \pm \nu \partial_{z} v = -2\nu^{2}\partial_{z}^{2}
\log{\tau_{0,1}}}
respectively and are square roots of partition functions of open-closed
string theory in the $(2,2m-1)$ minimal model backgrounds.
The unitary matrix partition function itself \unit\ is the product $\tau_{0}
\tau_{1}$ and so using the fact that $\tau_{0}$ and $\tau_{1}$ have open
string worldsheet
expansions with $\G = \pm 1/2$ respectively $(C=0)$, the continuum worldsheet
expansion of \unit\ has
only even numbers of
holes and $\G^{2}=1/4$. Although it might seem
odd that this picture of the
continuum limit is not all that similar to that of the $1/N$ expansion
before double-scaling, one must remember that some other matrix model with
`two-cut' behaviour of the eigenvalue density would have sufficed equally
\noflow\ .
Rather it should caution us that the universal surfaces
creamed-off by
the double-scaling limit need bear little resemblance to the finite ones we
see in the Feynman diagrams. It has been suggested that mKdV describes dense
polymer trees \ref\poly{{\v C}.Crnkovi\'{c}, M.Douglas,
G.Moore, Yale/Rutgers preprint
YCTP-P25-91 / RU-91-36.}.
Though such polymers have a `dual' description as dense self-avoiding loops
which is reminiscent of the holes created by open strings, especially
in view of the fact that
there is no boundary cosmological constant to suppress their length, we
have been unable to make a precise identification. Moreover the authors of
ref.\poly\ were considering the full 2nd mKP hierarchy, including the `even'
time variables. We have been working here with the reduced
case obtained by setting these to zero and using only the `odd' time
variables $t_{k}$.

Before closing this section let us mention the Virasoro constraints of
open-closed string theory on the tau-functions $\tau_{0}$, $\tau_{1}$. These
are the usual ones of closed string
theory except that $L_{-1}$ is {\sl apparently} missing and
$L_{0}$ contains an eigenvalue $\mu$ (as well as the $1/16$)
i.e.
$(L_n-\mu_i\delta_{n,0})\tau_i=0$ where $i=0,1$ and $n\ge0$
\TimH . Indeed, substituting $v=\partial_z \log (
\tau_1/\tau_0)$ into the mKdV flows and using eqn.\pII , one
readily deduces\foot{We thank T.Hollowood for bringing
this to our attention \ref\thanks{T.Hollowood, private
communication (July 1992).}.} that $\mu_1-\mu_0=C/2$.
Note also that the ${L}_{0}$ equation is the RG equation for the
$\tau$ function and is equivalent to \smiley\ for solutions of the KdV
hierarchy.
As
remarked previously, for open strings we are free to
 introduce a
constant $\rho$ which weights holes of length $l$ in the worldsheet by
${\rm e}^{-\rho l}$. For simplicity the discussion which follows is
understood to be perturbative\foot{Amongst other things this
obviates the need to burden the reader with certain non-perturbative
effects on the boundary cosmological constant which arise for DJM solution
(see \pqmodels\ref\simon{S.Dalley, Mod. Phys. Lett. {\bf A7} (1992) 1263.}
for details).}\ in $\nu$. We can generate
$\tau (\rho)$ from
$\tau (0)$ by using the fact that a boundary cosmological constant couples
to $L_{-1}$ \ref\bound{E.Martinec, G.Moore and N.Seiberg, \PL{B263}
(1991) 190.}; $L_{-1}\tau(\rho)=\partial
 \tau(\rho)/\partial\rho$. This equation may be confirmed
by a straightforward matrix-model calculation which
determines the r.h.s. as
\eqn\resolv{{\partial u \over \partial \rho}  = 2\G \nu\partial_{z}
{\rm diag}\{(-\nu^{2}\partial_{z}^{2} +u+\rho)^{-1}\}\ \ \ .}
It is then equivalent to $\R = -2\G \nu {\rm
 diag}\{(-\nu^{2}\partial_{z}^{2} +u+\rho)^{-1}\}$,
which is always true to all orders in perturbation theory.
The general solution is
$\tau (\rho) = {\rm e}^{\rho L_{-1}} \tau (0)$. This filters through to the
Virasoro constraints as
$$L_n\tau(\rho)={\rm e}^{\rho L_{-1}} \{
{1\over n!}(-\rho\ {\rm ad} L_{-1})^n\cdot L_n +
{1\over (n+1)!}(-\rho\ {\rm ad} L_{-1})^{n+1}\cdot L_n
\} \tau(0)$$
where ${\rm ad}L_{-1}\cdot L_m \equiv [L_{-1},L_m]$,
which on rearranging gives
\eqn\viras{{\tilde L}_{n} \tau(\rho) \equiv
\left[ L_n-(n+1)\mu\rho^n-
 \rho^{n+1} {\partial \over \partial \rho}\right]
\tau(\rho)=0 \hskip 1cm \forall n\ge -1 \ \ .
}
It is easy to confirm that these ${\tilde L}_n : n\ge-1$
satisfy the centreless Virasoro algebra.
We do not wish to dwell
upon the details here but merely emphasise the obvious but important fact that
\eqn\obvious{\G \neq 0 \Rightarrow {\partial \tau \over \partial \rho} \neq 0
 \ \ {\rm for\ \ any} \ \  \rho}
Thus when $\rho=0$, as we have been considering previously, the constraints
at $n \geq 0$ are the previous ones, while $n=-1$ remains there but
modified so long as we are dealing with open strings. Therefore $L_{-1}$ is
not missing but rather is neutralised as a constraint by the presence of
an extra variable other than the $t_{k}$.

\sec{Unitary Matrix Model in an External Field.}
In this final section we describe the equivalence  of the DJM closed
string solution $(\G=0)$ to eqn.\smiley\ and that
 of the $N\times N$
unitary matrix model  in  an external field and its formal multi-critical
generalisations  considered by  Gross and
Newman\gni\gnii . This involves the following integral
\eqn\extern{Z= \int {\cal D}U {\rm exp}(N{\rm Tr}
[UA^{\dagger} + U^{\dagger}A])}
which can arise in the mean field approach to lattice gauge theory.
$A$ is a complex `background' matrix and some Dyson-Schwinger equations for
$Z$ were used in \gni\ to solve the model in principle
to all orders in $1/N$, the model
depending only on the (positive) eigenvalues\foot{It seems likely that they
are related in some way to the eigenvalue path integral representation of
the DJM solutions \ref\complex{T.R.Morris, \nuke\  {\bf B356}
(1991) 703.}\multi\ .}\ $\lambda_{a}$ of $A^{\dagger}A$.
The authors  noted that  the critical  behaviour is  expressed
in  terms  of  the  (unscaled)
quantities  $\sigma_k={1\over N}\sum_b  \lambda_b^{-k/2}$ for  $k=1,3,\cdots$.
A special  r\^ole is  played by  $\sigma_1$ since  the model has two
phases `strong' and  `weak' according to $\sigma_1>2$ or  $<2$, the two phases
being split by a third order  phase transition.
Recalling the asymptotic expansions \posex\negex\ of the DJM solution,
precisely this order separates
the $z>0$ and $z<0$ spherical free energy, and suggests the
identification  $z\sim  \sigma_1-2$,  the  latter  having  been  identified in
ref.\gni\  as  proportional  to   the  `cosmological  constant'.  The  authors
recognized that the simplest critical point  in their model is essentially
the $m=1$ point of the ordinary  unitary matrix model \unit\ , which is known
to be true also for the $m=1$ DJM solution \multi\ref\early{T.R.Morris,
FERMILAB--PUB--90/136--T,
 to appear in
Jour. of Class. and Quant. Gravity}
quite apart from the Miura map.
This provides some intuition
for the coincidence. More immediately it  fixes the sign of the identification
between $z$  and their cosmological  constant to be  as above. Note  that this
implies that $z\to+\infty$ is their  strong coupling regime.
The multicritical
points in  ref.\gni\ are reached  by formally setting
$\sigma_3=\cdots=\sigma_{2m-1}=0$; in addition  in the continuum limit all
$\sigma_{2n+1}$ with $n>m$ scale to zero.  The authors find that there are
contributions to all genus in the strong  phase, while in the
 weak phase there
are contributions  only at genus  $g=1$~mod~$m$. This  agrees
with the DJM solution\multi\
  on specializing to an $m$-critical point
$t_k\propto   \delta_{k,m}$ (cf.eqns.\posex\negex\ with $\G=0$).  This
suggests   the  identification  $t_k\sim
\sigma_{2k+1}-2\delta_{k,0}$  up   to  scaling  which,   given  the  form   of
$\sigma_{2k+1}$  in terms  of $\lambda_b$,  is nothing  but a  Kontsevich-Miwa
transformation\ref\miwa{A.   Marshakov,  A.   Mironov,  A.   Morozov,
\PL{B274} (1992) 280.}\gnii  . This
transformation was performed for the unitary matrix model in ref.\gnii\ and we
refer  the  reader  there  for  details.   Indeed  if  we  set  $t_k  ={N\over
2k+1}(\sigma_{2k+1}-2\delta_{k,0})$  then from  eqn.(3.7) of  ref.\gnii\ it is
clear  that  their  Dyson--Schwinger  equations  are  precisely  those  of the
DJM solution\npqg  : the  Virasoro  constraints
  $L_n\tau=0$  for  $n\ge0$.  (The $L_{-1}$ constraint is tied up in a similar
way to that described in
the previous section, due this time not to open strings but to a purely
non-perturbative phenomenon which is described further in
refs.\pqmodels\simon\ref\talk{S.Dalley, C.V.Johnson and T.R.Morris, Nucl. Phys.
{\bf B} (Proc. Suppl.) {\bf 25A} (1992) 87, Proceedings of the workshop on
{\it Random Surfaces
and 2D Quantum Gravity}, Barcelona 10-14 June 1991.}.
The non-perturbative parameter
$\sigma$ which governs the behaviour, corresponding to an eigenvalue
representation of the DJM solution
on $[\sigma ,\infty)$,  is zero in the present case
(cf. footnote 6).
The authors of ref.\gnii\  do not in fact take the  continuum limit
but rather  analyse these equations  around a topological  point. This is  not
necessary. Our equations of course refer to the case where the continuum limit
has  already  been  taken\multi\  but  this  does  not  alter  the form of the
constraints since  they and the  $t_k$'s scale homogeneously  in the continuum
limit.)  Finally M.Newman has compared the
first few coefficients of the expansions for the multi-critical points of the
two theories and found precise agreement \ref\priv{M.Newman, private
communication (February 1992).}. Given all the asymptotic
coincidences,
the matching Virasoro constraints are persuasive
evidence also for non-perturbative agreement, however to
completely establish
equivalence
we would need to show for example that the solutions satisfy
 the same differential equation(s),
together with uniqueness properties given boundary conditions.

Accepting the equivalence, Gross and Newman's model
suggests
some interesting interpretations for the DJM solutions.
We referred above to the spherical level third order phase transition
at $z=0$. However  the full {\sl \nonp\ }  solution \npqg\ has no discontinuous
transition so  that it is  in reality at least an  infinite order
transition, if it  is correct  to  think   of  this  as
a  phase  transition  at
all. Similar  comments have been made in  ref.\ref\stoch{J. Ambj\o  rn, C.V.
Johnson and T.R.  Morris, Nucl. Phys. B374 [FS] (1992)  496.}.
Rather it has some similarity to a
`roughening' transition for the strong coupling phase world-sheet expansion,
a concept familiar  in lattice strong
coupling perturbation theory\ref\rough{C.
Itzykson,  M.E. Peskin  and J.B.  Zuber, Phys.  Lett. 95B  (1980) 259\semi  A.
Hasenfratz,  E.  Hasenfratz  and  P.  Hasenfratz,  Nucl.Phys. B180[FS2] (1981)
353\semi M. L\"uscher, G. M\"unster  and P. Weisz, Nucl.Phys. B180[FS2] (1981)
1.}. We  may think of  the  point  $z=0$ as being  the point where the
string fluctuations become  so large that the concept  of these strings breaks
down: all world-sheet observables are singular there although the exact theory
suffers no phase transition. One might wonder if there
is nevertheless some sort of string description in the weak
phase.  Interestingly this question may be answered by
 the $m=0$ `topological' point,
 a well-defined
expansion about a free integral over $U(N)$ identified in ref.\gnii . It
corresponds  to setting all $t_k=0$ except $t_0$,
whence the string equation\foot{The new topological point is
 simply inconsistent with $\R=0$.}\ \smiley\ (with $\G=0$)
yields the  expected trivial result $u=-1/(4z^2)$.
It follows that in general the $t_k\ne0$ `topological expansion'
 about this point  corresponds to the
 $z\to-\infty$ asymptotic expansion
of \smiley\ (e.g. \negex\ with $\G=0$).

\sec{Summary.}
We have shown how the string-theoretic solutions of the mKdV and KdV hierarchy
are unified by the Miura transformation in the general picture of open-closed
string theory, establishing connections between the non-singular solutions
of each. It would be interesting to know whether this geometrical picture
extends  more generally to the  KP and mKP hierarchy. We have also presented
overwhelming evidence that the unitary matrix model in a tuned external field
and its formal multi-critical generalisations
matches the known non-singular closed string solutions
of the KdV hierarchy. It seems likely that lurking behind the results
of this letter is a more general setting in terms of quantum gravity,
topological gravity and integrable hierarchies.

\bigskip\sec{Acknowledgements}
\noindent C.J. thanks the S.E.R.C. for financial support.
T.R.M. thanks Tim Hollowood for some enlightening correspondence.
S.D. acknowledges
discussions with Mike Newman and Andrea Pasquinucci, comments from D.Gross, and
is supported by
S.E.R.C. post-doctoral fellowship RFO/B/91/9033.

\listrefs
\bye